%% file: template.tex
\documentclass{vgtc}                          

\ifpdf
  \pdfoutput=1\relax                   
  \usepackage{graphicx}                
  \DeclareGraphicsExtensions{.pdf,.png,.jpg,.jpeg} 
\else
  \usepackage{graphicx}                
  \DeclareGraphicsExtensions{.eps}     
\fi%

\graphicspath{{figures/}{pictures/}{images/}{./}} 

\usepackage{microtype}                 
\PassOptionsToPackage{warn}{textcomp}  
\usepackage{textcomp}                  
\usepackage{mathptmx}                  
\usepackage{times}                     
\usepackage{cite}                      
\usepackage{booktabs}                  
\usepackage{tabularx}
\usepackage{threeparttable}
\usepackage{enumitem}
\usepackage{amsmath}
\AtBeginDocument{%
  \RequirePackage{cleveref}
  \crefname{figure}{Fig.}{Figs.}%
  \Crefname{figure}{Fig.}{Figs.}%
  \crefname{section}{Sec.}{Secs.}%
  \Crefname{section}{Sec.}{Secs.}%
  \crefname{table}{Tab.}{Tabs.}%
  \Crefname{table}{Tab.}{Tabs.}%
}

\usepackage{soul}

\newcommand{\revised}[1]{{\color{black} #1}}
\newcommand{\final}[1]{{\color{black} #1}}

\onlineid{9367}

\vgtccategory{Research}

\vgtcinsertpkg

\title{NotebookRAG: Retrieving Multiple Notebooks to Augment the Generation of EDA Notebooks for Crowd-Wisdom}

\author{
Yi Shan\textsuperscript{1}\thanks{E-mail: ydan24@m.fudan.edu.cn} \and
Yixuan He\textsuperscript{1} \and
Zekai Shao\textsuperscript{1} \and
Kai Xu\textsuperscript{2}\thanks{E-mail: kai.xu@nottingham.ac.uk} \and
Siming Chen\textsuperscript{1}\thanks{E-mail: simingchen@fudan.edu.cn. Corresponding author.} \\
}

\affiliation{
\textsuperscript{1}Fudan University, China\\
\textsuperscript{2}University of Nottingham, UK
}

\abstract{
   \input{./sections/0-abstract}
}

\begin{document}

\maketitle
\input{./sections/1-introduction}
\input{./sections/2-relatedwork}
\input{./sections/3-formativestudy}
\input{./sections/4-pipeline}
\input{./sections/5-casestudy}
\input{./sections/6-evaluation}
\input{./sections/7-discussion}

\acknowledgments{
This work was supported by the Natural Science Foundation of China (NSFC No. 62472099).
}

\bibliographystyle{abbrv-doi}
\bibliography{assets/bibs/papers}

\end{document}

%% file: sections/0-abstract.tex
High-quality exploratory data analysis (EDA) is essential in the data science pipeline, but remains highly dependent on analysts’ expertise and effort.
While recent LLM-based approaches partially reduce this burden, they struggle to generate effective analysis plans and appropriate insights and visualizations when user intent is abstract.
Meanwhile, a vast collection of analysis notebooks produced across platforms and organizations contains rich analytical knowledge that can potentially guide automated EDA.
Retrieval-augmented generation (RAG) provides a natural way to leverage such corpora, but general methods often treat notebooks as static documents and fail to fully exploit their potential knowledge for automating EDA.
To address these limitations, we propose NotebookRAG, a method that takes user intent, datasets, and existing notebooks as input to retrieve, enhance, and reuse relevant notebook content for automated EDA generation.
For retrieval, we transform code cells into context-enriched executable components, which improve retrieval quality and enable rerun with new data to generate updated visualizations and reliable insights.
For generation, an agent leverages enhanced retrieval content to construct effective EDA plans, derive insights, and produce appropriate visualizations.
Evidence from a user study with 24 participants confirms the superiority of our method in producing high-quality and intent-aligned EDA notebooks. 

%% file: sections/1-introduction.tex
\section{Introduction}

Exploratory Data Analysis (EDA)\cite{tukey1977exploratory} plays a crucial role in the data science pipeline.
High-quality EDA is time-consuming and requires coding proficiency, statistical thinking, and visualization literacy\cite{wongsuphasawat2019goals}, which places a heavy burden on analysts.
To ease these issues, prior research has explored rule-based and reinforcement learning approaches for automated EDA \cite{wu2024autoeda, bar2020automatically, yan2020auto}. With the rise of large language models (LLMs), automated EDA has gained powerful capabilities, enabling better alignment with user intent \cite{ma2023demonstration, zhu2024towards} and comprehensive insight discovery\cite{manatkar2024quis, zhao2024lightva}.
However, some approaches still face challenges in handling abstract intent, others may rely on comprehensive fact-checking of the dataset that can be inefficient and fragmented, and most remain limited in providing visualizations that effectively support analytical reasoning \cite{chen2025interchat, hutchinson2024llm}.

In real-world data mining, the process is typically driven by a high-level predictive or descriptive goal \cite{myatt2009making}. While analysts can usually specify the overall objective at the outset (e.g., building a time-series model for price prediction), 
\revised{they often struggle to design targeted EDA plans that effectively support and operationalize the data mining task}
(e.g., first providing an overview of the price series and then analyzing its seasonal distribution). 
This reflects an abstract intent \revised{(performing EDA to prepare for a data mining task)} that lies between explicit instructions (e.g., testing for significant periodicity) and no intent (e.g., simply asking to understand the data) \cite{wongsuphasawat2019goals}.

\revised{In practice, such as in Kaggle competitions or enterprise analytics, the same dataset or data source is often explored repeatedly by many analysts, producing numerous computational notebooks \cite{perkel2018jupyter} that share consistent data semantics and thus provide more relevant and reliable analytical knowledge for EDA \cite{rule2018exploration, deutch2020explained}.}
Interviews with senior enterprise analysts further confirmed this practice, revealing that analysts routinely revisit updated datasets and rely on existing notebooks to accelerate analysis and improve efficiency.

Retrieval-augmented generation (RAG) \cite{lewis2020retrieval} provides a natural way to leverage such notebooks, but existing approaches face two main issues. 
For retrieval, notebooks are often treated as static documents rather than executable artifacts, causing retrieved content to become invalid as data evolves; additionally, cells are handled independently, which ignores contextual dependencies and degrades retrieval quality \cite{li2024unlocking, li2023edassistant, li2021nbsearch}.
For generation, prior methods mainly target well-defined question answering \cite{tufino2025notebooklm} and are therefore ill-suited for automating EDA driven by abstract user intent.

To better leverage these corpora for automated EDA, we propose a method named NotebookRAG, which takes user intent, a dataset, and existing notebooks as input, first retrieving and enhancing relevant content from notebooks and then automatically generating EDA notebooks \revised{that integrate statistical analysis and visualization}.
For retrieval, code cells are enriched with contextual information, transformed into executable components, and annotated with the data columns they use.
The user intent (e.g., conducting EDA to prepare for time-series modeling) is mapped into multiple EDA queries (e.g., examining average price by region), which are used to retrieve relevant components based on their used columns. 
These components are re-executed on new data to generate updated visualizations, from which reliable insights are obtained.
For generation, we design an agent that produces EDA notebooks from the dataset and user intent, with an optional interface to incorporate retrieval outputs. 
Leveraging the enhanced retrieval content, the agent can construct more effective EDA plans, generate more appropriate visualizations, and derive insights that better support analytical reasoning.

To evaluate NotebookRAG, we conducted a within-subject user study with 24 participants using realistic Kaggle datasets, representative data mining tasks, and existing notebooks. Participants compared the quality of notebooks produced by the ChatGPT Data Analyst plugin \cite{chatgpt_data_analyst}, a baseline notebook generator, a general retrieval method \cite{li2024unlocking}, and our proposed retrieval method.
In addition, we performed objective checks on notebooks generated by our method. NotebookRAG was rated significantly higher than the other approaches across most evaluation dimensions and received more positive qualitative feedback, demonstrating its ability to generate higher-quality notebooks that better align with user intent.

In summary, our contributions are concluded as follows: 
\begin{itemize}[leftmargin=*,itemsep=0pt,topsep=0pt,parsep=0pt]
\item NotebookRAG, an automatic approach for generating efficient and effective EDA notebooks by combining user intent, datasets, and existing notebooks.
\item A retrieval technique that extracts relevant content from existing notebooks and an agent that leverages this content to automatically generate EDA notebooks.
\item A user study demonstrating that NotebookRAG significantly outperforms baselines in generating higher-quality, intent-aligned EDA notebooks.
\end{itemize}

%% file: sections/2-relatedwork.tex
\section{Related Work}

In this section, we review computational notebooks, automating EDA, and insight generation.

\subsection{Computational notebook}

Computational notebooks have become widely adopted for data analysis owing to their interactivity, reproducibility, and visualization capabilities\cite{rule2018exploration, perkel2018jupyter, kluyver2016jupyter, cardoso2019using}, leading to a vast number of notebooks hosted on public platforms such as GitHub \cite{huang2024contextualized, li2021nbsearch} and Kaggle \cite{mostafavi2024distilkaggle, quaranta2021kgtorrent}, as well as within enterprises\cite{deutch2020explained}.
By integrating executable code with rich documentation, notebooks serve both as computational environments and communication media; however, this dual role often leads to unstructured content, making notebooks harder to reuse and parse than traditional code files or analytical reports \cite{wang2020better, chattopadhyay2020s, wang2021makes, tian2025respark}.

\textbf{Notebook Reuse.}
Some works improve the understandability of personal notebooks to facilitate reuse by cleaning up messy content \cite{head2019managing} and enriching documentation with clearer explanations \cite{chattopadhyay2023make}.
Others have explored alternative presentation formats, such as slides \cite{wang2024outlinespark, wang2023slide4n}, reports \cite{wang2025jupybara}, or videos \cite{ouyang2024noteplayer} for storytelling, as well as visualizations of notebook content and structure \cite{wenskovitch2019albireo} to reduce the cognitive burden of reading notebooks. 
\revised{In addition, some studies have constructed large corpora of public notebooks \cite{mostafavi2024distilkaggle, quaranta2021kgtorrent}, 
which have been used to analyze real-world notebook practices \cite{quaranta2022assessing} 
by statistically characterizing various notebook features, 
and to support tasks such as code recommendation \cite{li2021nbsearch, li2023edassistant}, 
model training \cite{ghahfarokhi2024predicting}, and fine-tuning \cite{huang2024contextualized, li2025jupiter} 
by building mappings between code and natural language or by extracting task-specific code sequences.}

\revised{
Relatedly, ReSpark \cite{tian2025respark} extracts analytical objectives from existing reports and adapts them to new datasets, similarly emphasizing the reuse of analytical intent.
Improving the understandability of individual notebooks enables fine-grained local reuse, while corpus-based methods offer broader coverage; however, neither achieves both simultaneously.
In contrast, our work reuses multiple notebooks analyzing the same source datasets, combining fine-grained reuse with the diversity of analytical strategies across notebooks.
}

\textbf{Notebook Parsing.}
Notebook parsing is essential for understanding and analyzing notebook code, and existing methods generally fall into static and dynamic analysis.
Static analysis examines code via Abstract Syntax Trees (AST), commonly treating each code cell as a unit and representing it by the variables or APIs it uses, with relationships established across cells \cite{wenskovitch2019albireo, li2023edassistant, li2021nbsearch}.
Some approaches model the notebook as a linear yet segmented sequence under the assumption of sequential cell execution \cite{yan2020auto, huang2024contextualized}, but they mainly focus on data wrangling and do not cover other EDA stages.
Dynamic analysis monitors the notebook kernel to collect runtime records and variable states, enabling richer analyses \cite{harrison2024jupyterlab, eckelt2024loops, head2019managing, xie2024waitgpt}. 
However, it relies on execution logs and is typically implemented as interactive plugins, making it unsuitable for already existing notebooks; re-executing notebooks to obtain such logs is also unreliable due to non-linear execution and low reproducibility \cite{pimentel2019large, chattopadhyay2020s}.
Therefore, we adopt static analysis and extend prior approaches.

\subsection{Automating EDA}

Recognizing that high-quality EDA typically requires substantial expertise from analysts, researchers have explored ways to reduce these requirements by developing EDA assistants or creating a fully automated EDA process.

\textbf{EDA assistants.}
EDA assistants serve to support analysts during user-led analyses. 
Some works reduce the analyst's burden through code recommendations\cite{li2023edassistant, li2021nbsearch} or visualization recommendations\cite{lee2021lux}, while others ease the cognitive load by visualizing the EDA process\cite{sarvghad2016visualizing, wenskovitch2019albireo}, which helps analysts conduct more effective EDA.

\textbf{Fully Automated EDA.}  
Fully automated EDA shifts the analyst’s role from active exploration to interpreting generated content, significantly reducing workload. 
Rule-based methods \cite{wu2024autoeda, yan2020auto} provide strong controllability but lack flexibility, while deep reinforcement learning methods \cite{milo2018deep, bar2020automatically, bar2019atena} demonstrate better generalization and are capable of generating end-to-end workflows. 
With the advances of LLMs, fully automated EDA has gained stronger capabilities: tools such as the ChatGPT Data Analyst plugin \cite{chatgpt_data_analyst} can now effectively automate the entire EDA pipeline, and recent works further enable better alignment with user intent\cite{ma2023demonstration, zhu2024towards} as well as more comprehensive insight discovery\cite{manatkar2024quis, zhao2024lightva}.
However, these approaches still face key limitations: they rely on explicit user intent, analyze entire datasets inefficiently, and fail to generate visualizations that effectively support reasoning \cite{chen2025interchat, hutchinson2024llm}.
Therefore, our work attempts to build an intelligent agent that leverages knowledge from existing notebooks to better handle these challenges.

\subsection{Insight Generation from Visualizations}
\textbf{Statistical Methods.}
Visualization insight generation is a key component of visualization recommendation and composition.
Before the emergence of LLMs, visualization insight generation primarily relied on statistical methods, which involved designing specific statistical schemes tailored to particular tasks and visualization types to identify significant insights\cite{tang2017extracting, ding2019quickinsights, deutch2020explained, wang2019datashot, deng2022dashbot}.
With the advent of LLMs and their code-generation capabilities, reliance on predefined statistical functions has been reduced, enabling more generalizable insight generation while alleviating LLM hallucination issues \cite{zhao2024leva, weng2024insightlens, wang2025chartinsighter}.

\textbf{VLM-based Methods.}  
Moreover, as VLMs have advanced in visual understanding, generating insights via visualization-to-natural-language (vis2nl) methods has gained increasing attention \cite{huang2024pixels}.  
Several studies have shown that when VLMs are provided with clearly labeled and standardized visualizations, they perform well on chart question tasks \cite{wu2024chartinsights}, chart captioning tasks \cite{lim2025chartcap}, and visualization literacy tasks \cite{shao2025language}.  
Recent work also shows that combining data with its visualization further enhances VLM performance on broader data analysis tasks \cite{li2025does}.  
However, hallucination remains a pervasive challenge, often leading to factual errors or ambiguous semantics \cite{islam2024large, augustin2025dash}.  
To mitigate this issue, prior studies have explored strategies such as converting charts into structured tables for consistency checking \cite{huang2023lvlms}, and constructing curated chart–caption datasets for fine-tuning \cite{lim2025chartcap}.

To obtain reliable insights from visualizations in notebooks, we design a hybrid approach that combines statistical methods and VLM-based methods, where insights are first extracted using VLMs and then verified and refined by LLM-generated statistical code.

%% file: sections/3-formativestudy.tex
\section{Formative Study}

Through interviews with four senior enterprise analysts (each with over 10 years of experience), we confirmed that such scenarios occur not only on public platforms but also in enterprise settings, where multiple notebooks exist for the same or closely related datasets, and analysts often consult them to guide their work. 
This reinforced our belief that RAG could enhance automated EDA. 
To specify the design requirements for a RAG-based pipeline, we conducted one-on-one interviews with 12 master’s and PhD students in data science who regularly use computational notebooks.
\revised{The study examined how they reuse notebooks and interact with automated EDA tools to identify key design requirements for integrating the two.}

\subsection{Procedure}

The session began by asking participants to recall their past experiences conducting data analysis with notebooks, particularly whether they reused notebooks created by others.
All participants confirmed doing so, noting that reuse significantly improves their efficiency.
We then provided them with a commonly used Kaggle dataset, a data mining task, and five highly upvoted notebooks, encouraging them to browse these notebooks to gain a targeted understanding of the data. 
Participants were asked to think aloud during this process, allowing us to observe how they used them. 
\revised{Next, we introduced the ChatGPT Data Analyst plugin \cite{chatgpt_data_analyst}, a representative generative tool capable of performing automated EDA tasks.}
Using the same dataset along with several preset prompts, participants experienced the process and results of using the plugin for exploratory data analysis and were invited to provide comments and critiques.
Building on these, we introduced the concept of RAG, positioning it as analogous to reusing prior analyses during generation.
Finally, we presented our hypothetical RAG-based pipeline for EDA notebook generation and engaged participants in identifying concrete design requirements.

\subsection{Design Requirements}
We summarized the interview recordings and confirmed the design requirements (\textbf{DRs}) with the participants as follows:

\textbf{DR1: Goal-Aligned Extraction and Enhancement.}
When observing how participants used notebooks, we found that without prior knowledge of the dataset, it was difficult for them to quickly distinguish relevant parts. Most reported that reading notebooks sequentially was time-consuming, and agreed that automatically filtering potentially useful content would be valuable.
\revised{Several participants (6/12) mentioned that well-documented markdown notes greatly improved their understanding.}
\revised{All participants acknowledged that when data became outdated, the factual results in notebooks lost their validity, leaving only the analytical strategies still applicable, which reduced the value of existing notebooks.}
\revised{These observations suggest that the retrieval stage should extract content aligned with users’ specific analytical goals and enhance it with new data to obtain updated results and corresponding explanations, thereby improving the relevance and interpretability of reused content.}

\textbf{DR2: Efficient and Effective Results.}  
Several participants \revised{(5/12)} pointed out that GPT’s outputs were “standard but lacked depth,” often limited to simple visualizations with minimal analytical insights, which made the initial results frequently unsatisfactory. Although GPT could be prompted to continue analyzing, this process was inefficient compared to directly obtaining richer visualizations that facilitate data understanding. These findings highlight the need for automatically generated EDA results that, within limited steps, provide a coherent, comprehensive, and in-depth analytical process and deliver valuable insights supported by appropriate visualizations.

\textbf{DR3: Flexibility and Robustness.}
Several participants \revised{(4/12)}  with prior RAG experience pointed out that one cannot always expect valuable information from retrieved notebooks. In particular, they noted that relevant notebooks may not exist, especially for private datasets, or that available notebooks may be of low quality. This highlights the need for a system that can flexibly leverage retrieval when available while remaining robust and producing reasonable outputs even when retrieval provides limited or no support.

\textbf{DR4: Seamless Code Reuse.}
A majority of participants \revised{(9/12)} noted that although GPT’s outputs revealed the code used to generate visualizations, the code could not be directly modified on the page. They also pointed out that some code blocks were incomplete, with parts of the context hidden in memory rather than explicitly provided, which made reuse inconvenient.
Most participants agreed that producing the final output as an executable notebook would be a more practical solution, as it allows free modification and continuation of analysis. Therefore, the generated EDA results should take the form of notebooks, enabling analysts to modify and re-execute independent code cells for further exploration or downstream tasks.

%% file: sections/4-pipeline.tex
\begin{figure*}[t]
  \centering
  \includegraphics[width=0.9\linewidth]{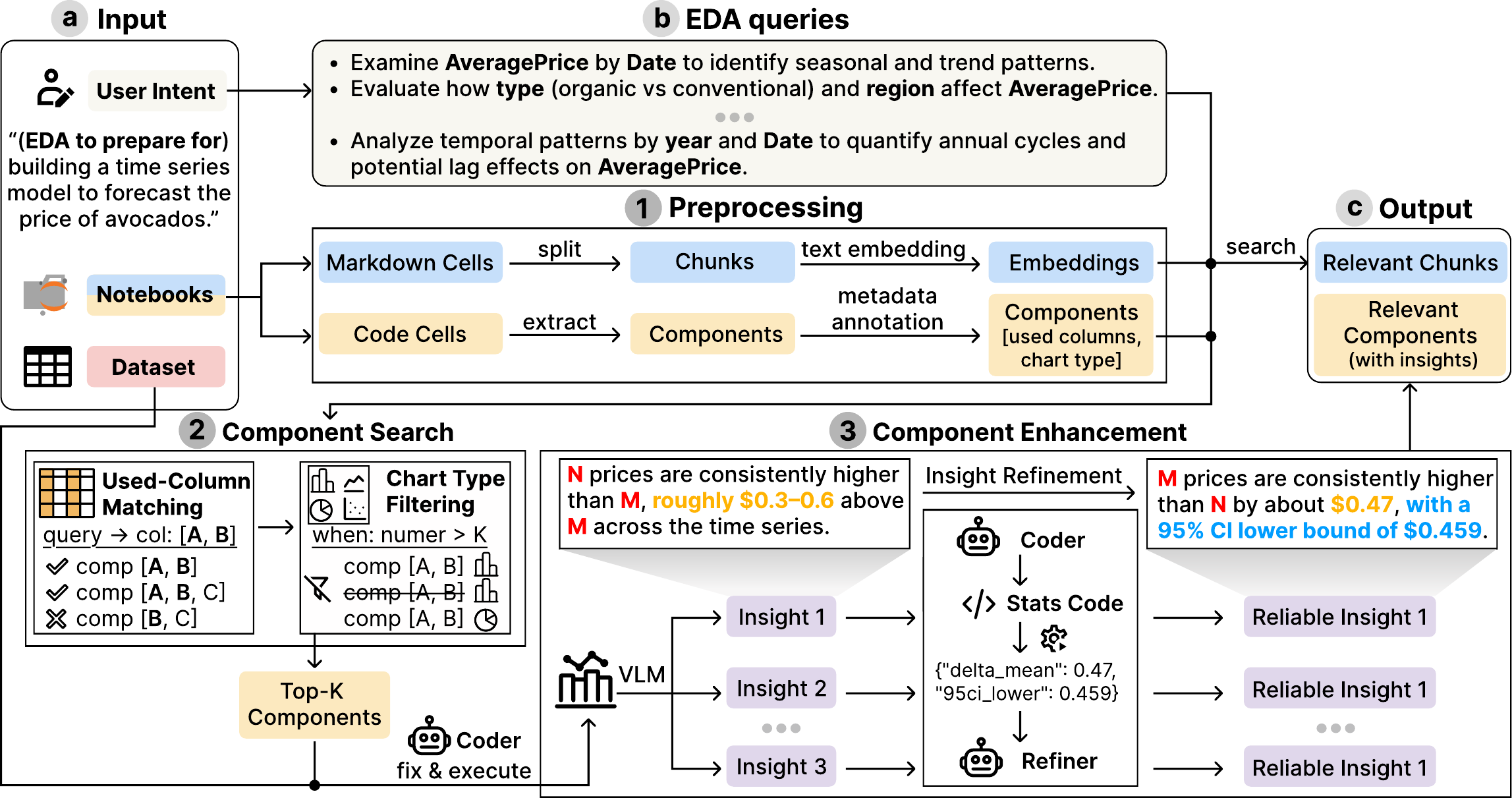}
  \caption{Overview of the Notebook Retrieval process. User intent (a) is mapped into EDA queries (b). Notebook markdown cells and code cells are preprocessed (1) into embeddings and components (with metadata), respectively. Then, the EDA queries are used to search embeddings for relevant chunks (c) and to guide component search (2) and enhancement (3), producing relevant components (c).
}
  \label{fig: retrieval}
  \vspace{-10px}
\end{figure*}

\section{NotebookRAG}

In this section, we first give an overview of NotebookRAG pipeline and two key steps: notebook retrieval (\Cref{fig: retrieval}) and generation (\Cref{fig: agent}).

\subsection{Pipeline Overview}

The pipeline takes as input (i) a tabular dataset, 
\revised{(ii) a collection of notebooks based on different versions of the same underlying data source, such as annual updates of a company’s revenue dataset,}
and 
\revised{(iii) a user intent, expressed in natural language, that specifies the subsequent data mining task (e.g., building a time-series model for price prediction), for which the system automatically constructs corresponding EDA notebooks as a preparatory step (as shown in \Cref{fig: retrieval}-a).}
The pipeline proceeds in two stages:

\textbf{Notebook Retrieval.}
Notebooks are first decomposed into code cells and markdown cells, since they serve distinct roles: code cells contain executable logic that produces data transformations and visualizations, while markdown cells capture human-authored explanations of analytical intent and conclusions. 
Code cells are transformed into executable components annotated with used columns and chart types, while markdown cells are converted into embeddings (\Cref{fig: retrieval}-1). 
\revised{The user intent is mapped into a set of EDA queries (\Cref{fig: retrieval}-b) based on LLMs, }
which are then used to retrieve relevant content: queries are matched with markdown embeddings for semantic similarity, their associated columns are used to search candidate components, and further filtered by chart type (\Cref{fig: retrieval}-2) (DR1).
Then, retrieved components are re-executed on the user-provided dataset, and a VLM extracts task-relevant insights from the resulting visualizations (\Cref{fig: retrieval}-3). These insights are then verified and refined with LLM-generated statistical code, which corrects factual errors and clarifies ambiguous statements to yield reliable insights (DR1).

\textbf{Notebook Generation. }  
We built an agent that automatically generates EDA notebooks from the dataset and user intent, with an optional interface to ingest retrieval outputs (DR3). 
The agent begins by constructing an EDA plan that specifies the goals and methods of each step, and then incrementally generates the corresponding visualizations and insights to produce a coherent, structured, and runnable notebook (DR4). 
Incorporating retrieved content enhances both the planning and generation stages, making the constructed plans more efficient and effective, the analysis code more in-depth, and the visualizations more appropriate (DR2).

\subsection{Notebook Retrieval}
\label{sec: notebook retrieval}

\subsubsection{Component Extraction}

\begin{figure}[t!]
  \centering
  \includegraphics[width=\linewidth]{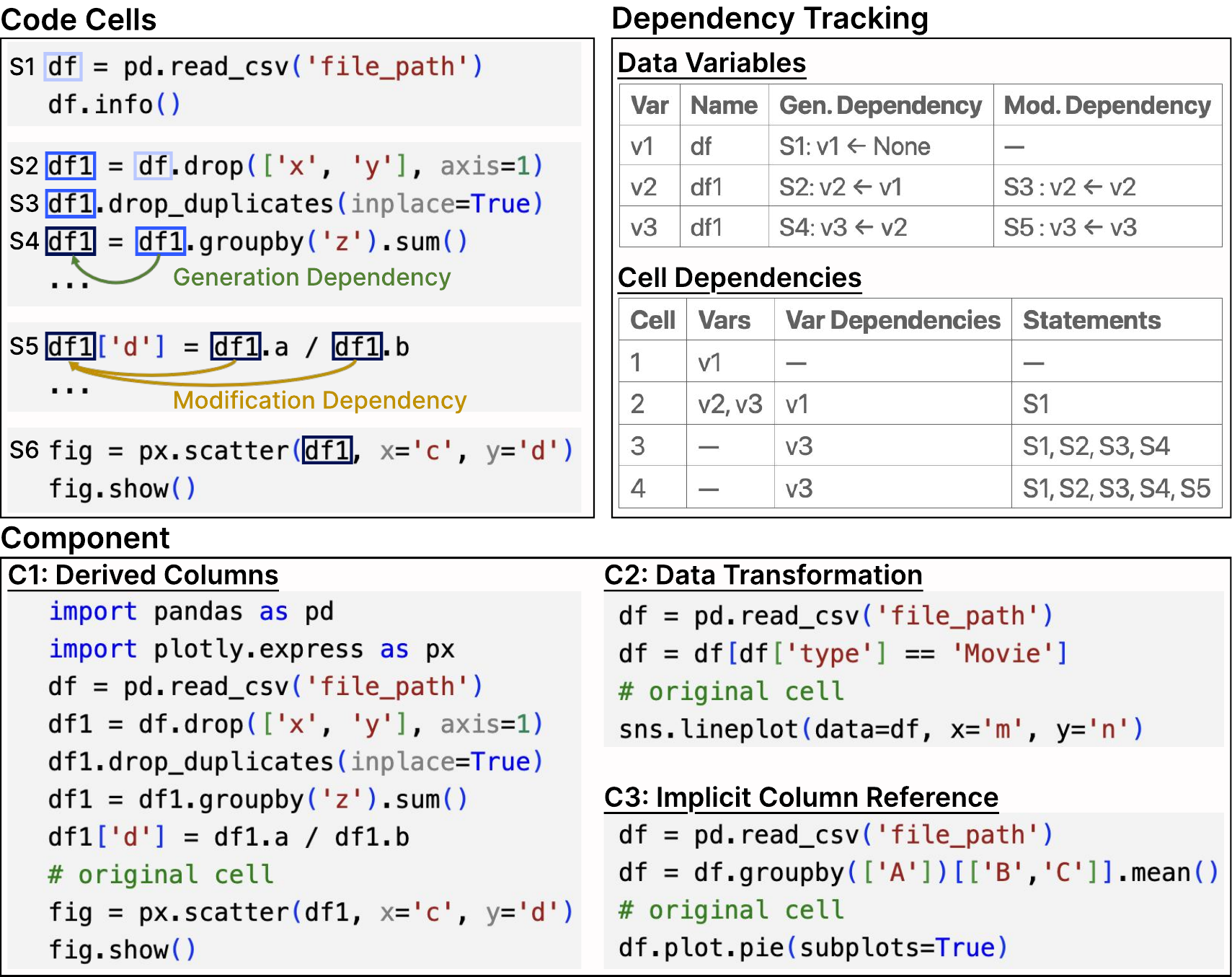}
  \caption{Example of the dependency tracking process. \textbf{Code Cells} define data variables through generation and modification statements. \textbf{Data Variables} summarize each variable’s lifecycle and dependencies. \textbf{Cell Dependencies} show how variables propagate across cells. \textbf{Component} collects the required statements into a self-contained unit that can be executed independently. Some cases where components improve annotation accuracy (C1, C2, C3).}
  \label{fig: component}
  \vspace{-20px}
\end{figure}

We define the \textbf{Component} {as shown in \Cref{fig: component}} as a self-contained, executable unit with resolved data and environment dependencies. 
Each notebook code cell that produces visualizations is transformed into such a component.
This design addresses two key challenges: (1) understanding the functionality of a single code cell often requires tracing implicit data dependencies and transformations; and (2) outdated dependencies and disorganized structure can easily lead to bugs, while debugging in a notebook environment is significantly less efficient than working with a continuous code block.
Considering the low reproducibility of public notebooks \cite{pimentel2019large}, we adopt a \textit{static analysis} approach rather than a dynamic one, thereby enabling a more general and robust method. 
Static analysis avoids execution-time failures caused by missing data or outdated dependencies, which are common issues in public notebooks. 
Our method supports both sequential and non-linear notebook executions, with the latter requiring a complete execution log. In the following, we describe our method assuming sequential execution.

We first merge code cells that generate visible outputs (e.g., plots or tables) or are followed by markdown cells with their preceding non-output cells.
To construct components, we introduce the \textbf{data variable}, which is any variable directly or indirectly derived from raw data. 
As illustrated in \Cref{fig: component}, data variables follow a lifecycle that includes generation (e.g., S1 creates \textit{df}, S2 and S4 create versions of \textit{df1}), modification (e.g., S3 and S5 update \textit{df1}), and downstream usage (e.g., S6 consumes \textit{df1} for visualization).
We then track these data-variable dependencies across the notebook in a top-down manner and resolve them at the cell level. 
Each statement that creates, modifies, or uses a data variable is identified, and its dependency chain is recursively traced. 
In this way, we record for every data variable the complete sequence of statements that describe its evolution from raw input to its current state (see Data Variables and Cell Dependencies in \Cref{fig: component}).
Finally, to construct the component, we prepend the minimal set of required statements for the data variables in a target code cell, preserving execution order to form a self-contained, executable unit.

In practice, we implement this process through an AST-based algorithm, which is robust to real-world complexities such as branches, loops, and function calls, and leverages a taxonomy of common data-processing libraries (e.g., \texttt{pandas}) to recognize implicit state mutations (e.g., method calls like \texttt{df.drop\_duplicates()}).

\subsubsection{Component Metadata Annotation}
Given an EDA query (e.g., “evaluate how type affects price”), semantic similarity search often performs poorly because of the semantic gap between natural language and code, the lack of method-level details, and the fact that many visualization functions (e.g., \texttt{sns.pairplot}) reference columns implicitly rather than explicitly \cite{gu2018deep, wu2024autoeda}.
To improve retrieval precision, we leverage the code understanding capabilities of LLMs to annotate each code snippet with the used columns and chart type (provided with dataset column descriptions).
\revised{Notably, the LLM-based annotation does not rely solely on explicit variable names. It infers column dependencies based on the semantic context of the code, such as understanding the effects of data transformations, thereby maintaining accurate mappings between queries and transformed variables (\Cref{fig: component}-Component).
}

To evaluate this approach, we manually annotated 840 pairs of code cells and corresponding components with metadata and compared the labeling accuracy of different models (\Cref{tab:usedcol_acc}).
For \textbf{chart type}, all models achieved high accuracy, with little difference between code cells and components. For \textbf{used columns}, however, we observed clear improvements when annotating components: across both SOTA models and smaller models, component-level annotation consistently outperformed cell-level annotation.
The advantage of components comes from preserving complete contextual information. 
We highlight some common cases: 
(1) the use of previously created derived columns (\Cref{fig: component}-C1), 
(2) data transformations applied upstream (\Cref{fig: component}-C2), and 
(3) visualization methods that implicitly reference columns without explicitly naming them (\Cref{fig: component}-C3).

\textit{Practical Note.} 
While \textbf{gpt-5-nano} achieved the best trade-off between accuracy and efficiency, 
\textbf{Qwen-7B} offers a practical alternative for large-scale annotation: when deployed on a single RTX 4090 GPU with 8 threads, it completed all 840 annotation tasks within one minute.

\begin{table}[t]
\centering
\footnotesize
\setlength{\tabcolsep}{4pt}
\caption{Accuracy of used-column annotation on code cell--component pairs. 
$\Delta$ (pp) denotes absolute percentage-point gain; $\Delta$ Rel. (\%) denotes relative improvement.}
\label{tab:usedcol_acc}
\begin{tabular}{lrrrr}
\toprule
\textbf{Model} & \textbf{Acc-Code (\%)} & \textbf{Acc-Comp (\%)} & $\boldsymbol{\Delta}$ \textbf{(pp)} & $\boldsymbol{\Delta}$ \textbf{Rel. (\%)} \\
\midrule
gpt-5      & 78.1 & 91.4 & 13.3 & 17.1 \\
gpt-5-nano     & 76.0 & 89.6          & 13.6          & 18.0 \\
Qwen-7B    & 63.5 & 74.0          & 10.5          & 16.7 \\
CodeL-7B   & 57.3 & 68.0          & 10.7          & 18.7 \\
DeepSeek-6.7B & 58.8 & 67.5       & 8.7           & 14.8 \\
\bottomrule
\end{tabular}

\vspace{2pt}
{\scriptsize \emph{Abbreviations:} 
Qwen-7B = \texttt{Qwen2.5-Coder-7B-Instruct}; 

CodeL-7B = \texttt{CodeLlama-7b-Instruct}; 

DeepSeek-6.7B = \texttt{Deepseek-Coder-6.7B-Instruct}.}
\vspace{-10px}
\end{table}

\subsubsection{Intent-Guided Retrieval}

\revised{Since user intent can often be abstract or ambiguous, we adopt an LLM-assisted exploratory strategy for intent interpretation and retrieval (as shown in \Cref{fig: retrieval}), inspired by question-guided insight generation \cite{manatkar2024quis}.
Specifically, we prompt LLMs with contextual information such as the dataset description to decompose the intent into a set of possible EDA queries (as shown in \Cref{fig: retrieval}-b), aiming to cover potentially valuable analytical sub-tasks to enhance the completeness and diversity of retrieval.}
Each query specifies target columns and analytical goals, which are then used to retrieve relevant content from both markdown cells and code components.
\revised{For markdown cells, considering the diversity of their content such as analytical objectives and conclusions as well as their unstructured narrative forms, we adopt a conventional approach that segments the text and performs embedding-based similarity retrieval to identify potentially relevant passages.
In particular, markdown content is excluded when data versions differ, which may lead to outdated conclusions.
}
For components, we leverage metadata to match the columns specified in the query: exact matches are prioritized, followed by partial matches that cover most of the query columns. If the number of candidates exceeds the top-$k$ (set to five), we further filter them by retaining only one component for each unique combination of used columns and chart type (as shown in \Cref{fig: retrieval}-2).

\subsubsection{Component Enhancement}
\label{sec: enhancement}

Since visualizations in the original notebooks may be invalid due to outdated data, and some components may contain obsolete code that prevents direct rerun, we enhance the retrieved components after top-$k$ selection (\Cref{fig: retrieval}-3). Specifically, we provide each component’s code together with the user’s dataset to a \textit{Coder}, which executes the code in a sandbox environment and automatically repairs errors, thereby producing runnable code and updated visualizations.

Given the strong performance of SOTA VLMs on various vis2nl tasks\cite{shao2025language, wu2024chartinsights, lim2025chartcap} and the demonstrated benefits of visual inputs for data analysis\cite{li2025does}, we rely on human-authored visualizations designed to support analytical reasoning to obtain insights that are difficult to derive from code or statistics alone.
We provide the user intent and the visualization images to a VLM, which produces actionable insights \cite{wang2025jupybara}, which reflect an understanding of the data and support informed decision-making in real-world contexts.

Considering that VLMs may produce hallucinations leading to factual errors in insights, or semantic ambiguities due to limited information in the visualization \cite{wang2025chartinsighter, huang2024pixels, yang2023dawn}, we introduce a refinement step. Specifically, the \textit{Coder} generates statistical code based on the content of the insight to compute precise data facts, and the \textit{Refiner} integrates the original insight with these results to correct errors and clarify ambiguities, ultimately producing reliable insights. 
For example, as shown in \Cref{fig: retrieval}-3, ``Insight 1'' initially contained a factual error (stating $N > M$ instead of $M > N$) and a vague description (“roughly \$0.36–0.6”). After refinement, the factual error was corrected ($M > N$), the ambiguity was resolved (\$0.47), and additional information was provided (a 95\% CI lower bound of 0.459).
\revised{A small-scale manual review further indicated that this refinement process effectively mitigates hallucinations by improving both factual accuracy and interpretive clarity.}

\subsection{Notebook Generation}

\begin{figure}[t!]
  \centering
  \includegraphics[width=\linewidth]{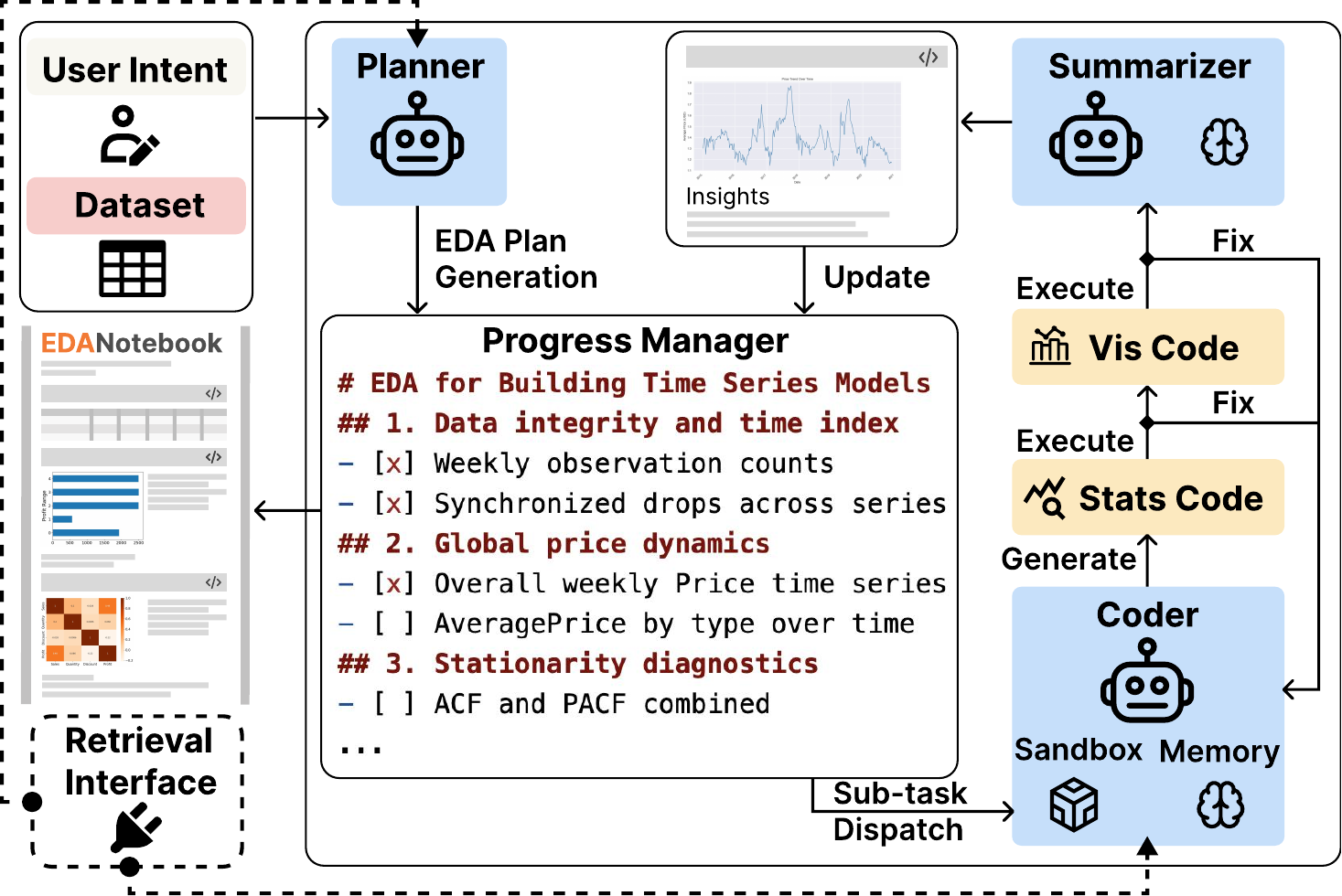}
  \caption{Agent for EDA notebook generation with retrieval interface. The agent takes user intent, a dataset, and retrieval outputs (optional) as input. It begins by constructing an EDA plan (Planner), then incrementally dispatches sub-tasks (Progress Manager), generates statistical code and visualization (Coder), and integrates results (Summarizer) to update the plan, ultimately producing a complete EDA notebook. Retrieval outputs are passed through a retrieval interface to enhance both planning and code generation.}
  \label{fig: agent}
  \vspace{-10px}
\end{figure}

As shown in Fig.~\ref{fig: agent}, we build an agent to automatically generate EDA notebooks, inspired by the design paradigm of intelligent agent widely adopted in prior work \cite{yao2023react, wang2023voyager, fourney2024magentic}, where a high-level plan is first generated, specialized tools are then invoked to complete sub-tasks, and global state is maintained through a centralized progress tracker to ensure coherent execution.

Given a dataset and user intent, the Planner formulates a natural language EDA plan specifying the goals and methods of each analytical step. 
The Progress Manager tracks and updates the plan to maintain coherence and assigns sub-tasks to the Coder.
The Coder first generates statistical code for the sub-task and executes it to obtain results, and then uses these results together with the sub-task descriptions to generate visualization code. 
All code is executed in a sandbox environment to ensure robustness and safety, and the Coder is equipped with memory to preserve continuity across code fragments generated for different sub-tasks.
The Summarizer, also equipped with memory, organizes the Coder’s outputs into units containing visualizations and actionable insights, and then submits them to the Progress Manager for updates. 
Once the plan is completed, the Progress Manager generates a final summary report together with recommendations for subsequent tasks and assembles all content into a coherent EDA notebook.

A key feature of our pipeline is the \textbf{Retrieval Interface}, which enriches the agent with reusable notebook components, reliable insights, and relevant markdown derived from the Notebook Retrieval stage (\Cref{sec: notebook retrieval}).
During plan construction, the Planner can ingest these retrieval outputs as additional context.
To support more general scenarios, we provide resources with descriptions and inform the Planner that they may refer to them, without explicitly instructing how to use.
This implicit guidance already yields substantial improvements in both the efficiency and effectiveness of the generated plans, as we demonstrate later in the case study (\Cref{sec: casestudy}).
During code generation, the Coder selectively reuses validated components based on their alignment with the sub-task goal and visualization suitability, while also using them as references when direct reuse is infeasible, producing visualizations that are closer to human-authored designs. 

To support traceability and source transparency, whenever components are reused, the generated notebook includes links at the corresponding positions that allow users to jump to the original notebook context. 
This enables analysts to inspect the source and discover potentially useful details, such as drill-down analyses. 

The retrieval interface is optional. 
When no existing notebooks are available, the agent simply proceeds without retrieval and can still generate complete EDA notebooks. 
When existing notebooks are available, the interface is always activated: high-quality notebooks enhance planning and generation, while low-quality or less relevant notebooks are filtered during the retrieval stage, thus having limited impact and ensuring the robustness of the overall pipeline.

\subsection{Implementation}

We use gpt-5-mini for extracting insights from visualizations and gpt-5-nano for text reasoning and code generation. 
\final{For semantic retrieval, notebook markdown cells are encoded using text-embedding-3-large and indexed with FAISS for similarity search.}
The overall workflow is orchestrated with LangChain, enabling modular agent construction and graph-based control of planning, execution, and summarization.

%% file: sections/5-casestudy.tex
\section{Case Study}
\label{sec: casestudy}

In this section, we qualitatively analyze the impact of incorporating retrieval on notebook generation. 
We first introduce a concrete usage scenario to set the stage, and then present several representative cases that highlight how retrieval affects both the overall EDA workflow and individual sub-tasks.

Consider an analyst aiming to build a time-series price prediction model on the Kaggle \textit{Avocado Prices (2020)} dataset.
Before modeling, the analyst wishes to explore the data by consulting existing notebooks. 
As this dataset is relatively new and has limited community analysis, whereas the earlier \textit{Avocado (2018)} dataset has been extensively studied, with many notebooks publicly available. 
Manually reviewing these notebooks is time-consuming, as the analyst must re-run them on the new dataset, debug any errors encountered, and then sift through large amounts of content to identify the parts that are actually useful.
With our system, the analyst only needs to download several highly upvoted notebooks from the old dataset and provide them with the new dataset and the user intent (``build a time-series model to forecast avocado prices''), as inputs. 
The system then generates EDA notebooks under two modes: with retrieval (RAG-enhanced agent) and without retrieval (baseline agent).
Incorporating retrieval content improved performance at both the global level (overall EDA plan) and the local level (individual sub-tasks), as illustrated by the representative cases below.

\begin{figure}[t!]
  \centering
  \includegraphics[width=\linewidth]{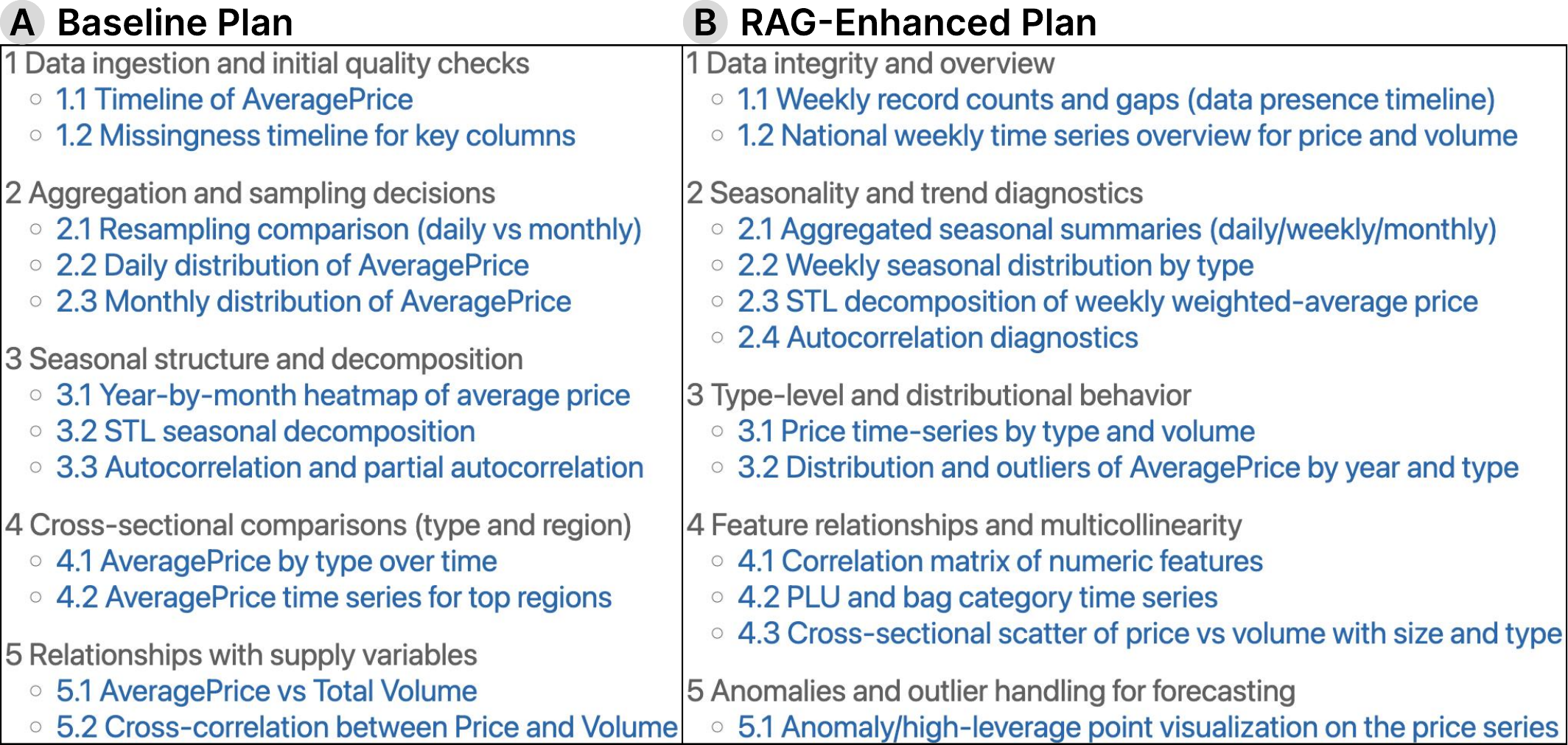}
  \vspace{-15px}
  \caption{A: Baseline EDA plan. B: RAG-enhanced EDA plan.}
  \label{fig: plan}
  \vspace{-15px}
\end{figure}

\textbf{Global Level.}
At the global level, the RAG-enhanced agent generated plans with clearer analytical priorities. 
Since LLMs operate as black boxes, our conclusions about how specific retrieval outputs influenced the final results are primarily derived from comparative analysis.
For example, when many retrieved insights indicated that “certain attributes exhibit clear annual seasonality,” the \textbf{RAG plan} (Fig.~\ref{fig: plan}) prioritized \textit{Seasonality and trend diagnostics} as an early step (Step~2), whereas the Baseline plan delayed this analysis until later (Step~3). 
By introducing this focus earlier, the RAG plan was able to perform a more fine-grained decomposition of seasonal and trend patterns within limited steps.
Similarly, when retrieved insights emphasized “anomalies” and “correlations,” the RAG plan explicitly dedicated sections to \textit{Feature relationships and multicollinearity} (Step~4) and \textit{Anomalies and outlier handling for forecasting} (Step~5). 
In contrast, the \textbf{baseline plan} lacked explicit anomaly handling and only included a simpler correlation analysis in terms of cross-correlation with supply variables. 
These enhancements in the RAG plan not only lead to a more structured and comprehensive workflow but also provide results that are directly useful for downstream tasks such as feature engineering in predictive modeling.

\begin{figure}[t!]
  \centering
  \vspace{-5px}
  \includegraphics[width=\linewidth]{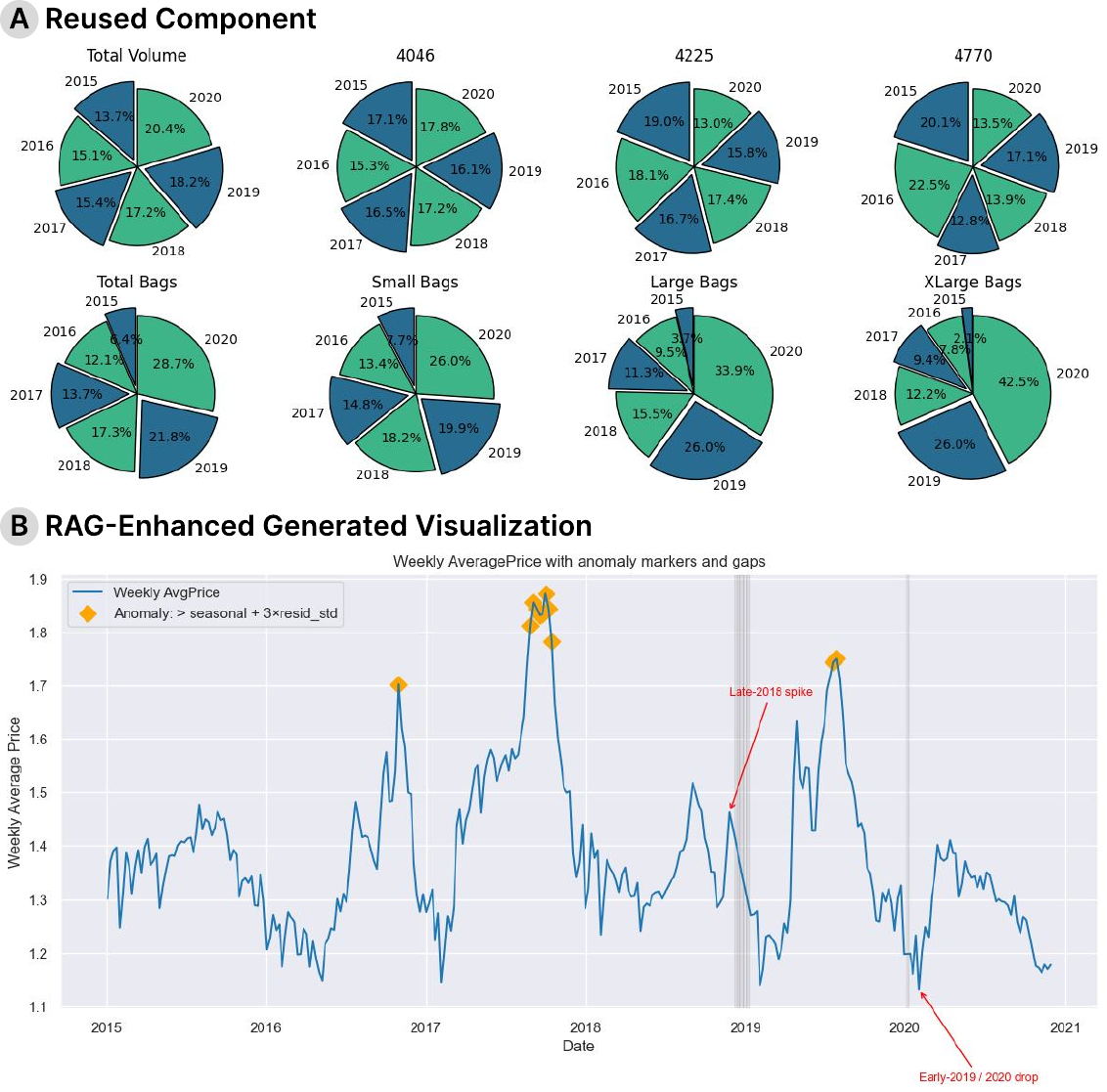}
  \vspace{-15px}
  \caption{A: a reused human-authored component showing the year-by-year distributions of features. B: a RAG-enhanced visualization highlighting anomalies and annotated events in the weekly average price series.}
  \label{fig: vis_example}
  \vspace{-15px}
\end{figure}

\textbf{Local Level.}
At the local level, the improvements are reflected in both the richness and appropriateness of visualization forms and the depth of analysis. 
Compared to the baseline, the RAG-enhanced agent produced more diverse outputs, such as maps, pie charts, and interactive visualizations. 
This diversity is enabled by the availability of reusable components. 
For example, at the beginning of the data overview, the Coder chose to reuse a component (\Cref{fig: vis_example}-A) that visualizes the \textit{year-by-year distribution across key features}. 
Through carefully designed, human-authored encodings, this visualization compactly presents multiple distributions in a limited space, allowing users to quickly obtain an overview of the dataset and facilitating more efficient exploration.
RAG-enhanced agent also adopted deeper analyses and more suitable presentation forms in certain cases. 
For instance, when insights highlighted anomalies such as “One pronounced anomaly (late-2018/early-2019): price spikes averaging 2.56 versus 1.38 during non-spike periods, occurring in low-volume periods” or seasonal fluctuations such as “AveragePrice shows seasonal fluctuations, rising to a peak price of 3.25 in 2016, then declining with a -0.02 delta from 2015 to 2018 and -0.07 from 2018 to 2020,” the Planner explicitly instructed the analysis of “gaps” and defined anomaly points as “weeks where price exceeds the seasonal component by more than 3 × residual\_std.” 
Based on these instructions, the Coder generated statistical code to identify anomalies, and then used these results to produce visualization code that not only marked the anomaly points but also incorporated appropriate annotations, highlighting events such as the late-2018 price spike and the early-2019/2020 price drop (\Cref{fig: vis_example}-B). 
\final{The system also provided actionable insights, such as “missing-week gaps: enforce a continuous weekly index and impute with seasonally-aware methods before ARIMA/LSTM,” which guided the subsequent analysis and suggested improvements for modeling.}
This integration of statistical analysis and visual annotation provides a clearer and more informative representation of the data patterns.

%% file: sections/6-evaluation.tex
\section{User Study}

\subsection{Datasets \& Materials} 

To ensure that our user study closely reflects real-world scenarios, we carefully prepared the experimental materials along three dimensions: datasets, tasks, and notebooks.  

\textbf{Datasets.} 
We selected three commonly analyzed Kaggle datasets that represent typical scenarios frequently encountered by analysts: \textit{Avocado Prices} (continuously updated), \textit{Superstore} (available in many versions), and \textit{Netflix Movies and TV Shows} (continuously updated). These datasets were chosen because they not only appear repeatedly in public analysis platforms but also mirror situations where similar or updated datasets arise in enterprise settings.  

\textbf{Tasks.} 
\revised{We provided participants with a description of a data mining task as their user intent and asked them to evaluate whether the generated EDA notebooks could help them better prepare for the task.
Because participants might not be familiar with the datasets, the task was predefined rather than self-formulated, ensuring comparable analytical objectives and fairer cross-session evaluation of retrieval and generation outcomes.}
To capture different analytical goals, we included one predictive task and one descriptive task for each dataset. 
The tasks were selected based on the most common analyses observed on these datasets, ensuring that the retrieval mechanism would have relevant prior material to draw from. 

\textbf{Notebooks.} 
For each dataset, we collected the top 20 Python notebooks ranked by upvotes on Kaggle. 
This selection strategy aligns with the way users in real scenarios tend to prioritize higher-quality notebooks, while the number (20) is much larger than what would normally be examined manually, and also approximates the number of notebooks realistically available to analysts.

\final{Together, these choices align the study design with realistic analytical contexts while providing a controlled evaluation setting, using three datasets with two tasks each.}

\subsection{Study Design}

\begin{figure}[t!]
  \centering
  \includegraphics[width=\linewidth]{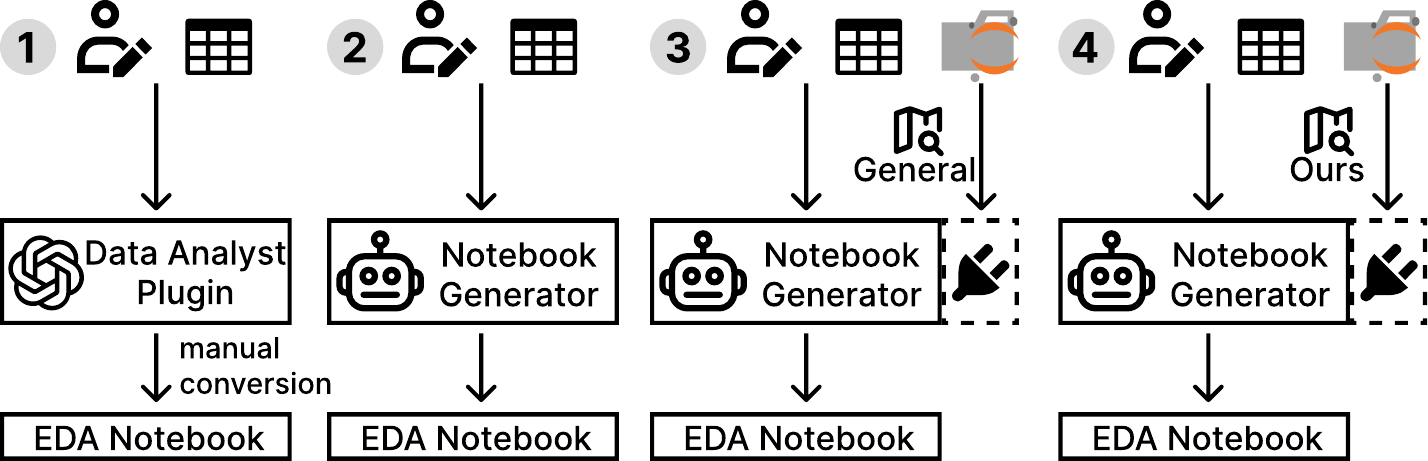}
  \caption{Overview of the study design. Four EDA notebooks were generated for evaluation: (1) using the ChatGPT Data Analyst plugin with manual conversion to an \texttt{ipynb} file, (2) using the baseline notebook generator without retrieval, (3) using the baseline generator with a general retrieval method, and (4) using the baseline generator with our proposed retrieval method.}
  \label{fig: ablation}
  \vspace{-10px}
\end{figure}

To systematically evaluate our approach, we generated notebooks for each dataset-task combination using four methods, with one notebook produced per method, as illustrated in Fig.~\ref{fig: ablation}.

\textbf{ChatGPT Data Analyst Plugin (\textit{ChatGPT}).} 
The first notebook was produced using the ChatGPT Data Analyst plugin. We provided the user intent (i.e., a data mining task with the explicit goal of performing EDA to gain targeted understanding) and the dataset, and asked the plugin to generate a complete EDA process. 
The generated process was then converted into an \texttt{ipynb} file, with minor manual corrections applied to ensure consistency with the origin and to match the format of notebooks generated by our method.

\textbf{Baseline Notebook Generator (\textit{Baseline}).} 
The second notebook was directly generated by our baseline notebook generator without using any retrieval, serving as the baseline for comparison.  

\textbf{General Retrieval (\textit{RAGBaseline}).} 
The third notebook incorporated a general retrieval method. Specifically, markdown cells and code cells from prior notebooks were separately embedded, and the user intent was used to perform semantic similarity search. The retrieved content was then passed into the retrieval interface, guiding the notebook generator to produce the final EDA notebook.  

\textbf{NotebookRAG (\textit{Ours}).} The fourth notebook used our proposed retrieval method, where retrieved components were passed into the retrieval interface and integrated into the generation process, enabling the system to reuse human-authored content and produce enhanced EDA notebooks.  

\textit{ChatGPT} is a well-established product and is used as a reference point without extensive prompting. 
\textit{Baseline} denotes our generator without retrieval, which, with engineering on gpt-5-nano, we estimate to be more powerful than \textit{ChatGPT} and thus serves as a stronger baseline and an ablation of our retrieval component. 
\textit{RAGBaseline} augments \textit{Baseline} with a general retrieval method, against which we compare \textit{Ours} to evaluate the effectiveness of our proposed retrieval approach. 
\revised{To ensure fair comparison, all methods were constrained by the same moderately relaxed step limit, enabling evaluation under comparable exploration budgets.}

\subsection{Methods}
We employed a within-subjects design to evaluate the notebooks produced by the four methods.
\revised{Unlike data analysis tasks that can be evaluated using objective metrics such as accuracy \cite{li2025jupiter}, objective measures like the correctness of generated visualizations \cite{helali2025reliable} cannot comprehensively assess the overall quality of EDA. 
Therefore, following most prior studies on automated EDA \cite{manatkar2024quis, ma2023demonstration, zhu2024towards}, we adopted human evaluation and further extended existing evaluation dimensions based on the identified design requirements.}
Participants received the materials and rated in a questionnaire consisting of 13 statements, as shown below, each corresponding to an evaluation dimension.
The quality sub-dimensions were designed by two external experts to ensure impartiality.
Responses were collected using a five-point Likert scale ranging from \textit{strongly disagree} to \textit{strongly agree}.
In addition, participants were encouraged to provide justifications or think-aloud comments for each rating.

\noindent \textbf{Overall Dimensions.} 
(1) \textit{\textbf{Confidence:}} I am confident in the validity and reliability of the analysis. 
(2) \textit{\textbf{Helpfulness:}} The EDA is helpful in exploring the data and supports me effectively in understanding it. 
(3) \textit{\textbf{Satisfaction:}} The analysis meets my expectations and leaves me satisfied with its overall quality and usefulness. 
(4) \textit{\textbf{Quality:}} N/A (This dimension is not measured by a single scale but is computed as the average of the ten quality sub-dimensions.)

\noindent \textbf{Quality Sub-dimensions.} 
(1) \textit{\textbf{Task Alignment:}} The analysis closely aligns with the stated data mining task and research goals. 
(2) \textit{\textbf{Data Comprehension:}} The notebook demonstrates a thorough understanding of the dataset, including missing values, outliers, and data quality issues. 
(3) \textit{\textbf{Coverage:}} The exploration covers a sufficient range of variables and their relationships (univariate, bivariate, multivariate). 
(4) \textit{\textbf{Visualization:}} The visualizations are appropriate, clearly presented, and helpfully support interpretation. 
(5) \textit{\textbf{Methodology:}} The statistical methods used are suitable, clearly explained, and properly interpreted. 
(6) \textit{\textbf{Insight:}} The notebook generates meaningful and non-trivial insights beyond simple descriptive summaries. 
(7) \textit{\textbf{Robustness:}} The analysis identifies and discusses potential biases, anomalies, or limitations in the data. 
(8) \textit{\textbf{Narrative:}} The narrative and explanations are coherent, logical, and easy to follow. 
(9) \textit{\textbf{Reproducibility:}} The notebook is reproducible, with clear code, documented steps, and environment/dependency specifications. 
(10) \textit{\textbf{Efficiency:}} The analysis is efficient and concise, avoiding redundancy while maximizing insight.

Evaluations were conducted under natural conditions, with a reasonable time limit of three days but without further restrictions or supervision.
To minimize bias, we employed a blinding procedure. 
Participants were unaware of the study’s purpose or our work and were instructed to evaluate only the four notebooks provided. 
The notebooks were formatted similarly (see supplementary material), preventing participants from inferring their origin and ensuring a fair comparison focused on quality.
Although the notebooks were anonymized, the order of presentation in the material folder could influence the sequence in which participants read them.
To control for order effects, we applied a balanced Latin square design to counterbalance the notebook order.
This design required at least four participants for each dataset-task combination.
Consequently, we recruited 24 participants across the six combinations in our study.

\final{To further validate the fine-grained performance of our approach, two co-authors conducted objective checks on the notebooks generated by \textit{Ours}, examining both the coverage of task-relevant key variables and the correctness of the generated analytical insights.}

\subsection{Participants}

Because the study focused on notebook quality, participants were required to have at least three years of data analysis experience and to have conducted analyses regularly (at least once per month in the recent past).
All participants were also required to be frequent computational notebook users.
Recruitment was conducted via peer recommendations on social media to ensure appropriate qualifications, which were further validated through examination of their qualitative feedback.
One participant exhibited weak reasoning, prompting us to recruit an additional participant as a replacement.
The final dataset consisted of evaluations from 24 participants.

\subsection{Results}
\label{sec: results}

\begin{figure*}[t]
  \centering
  \includegraphics[width=\linewidth]{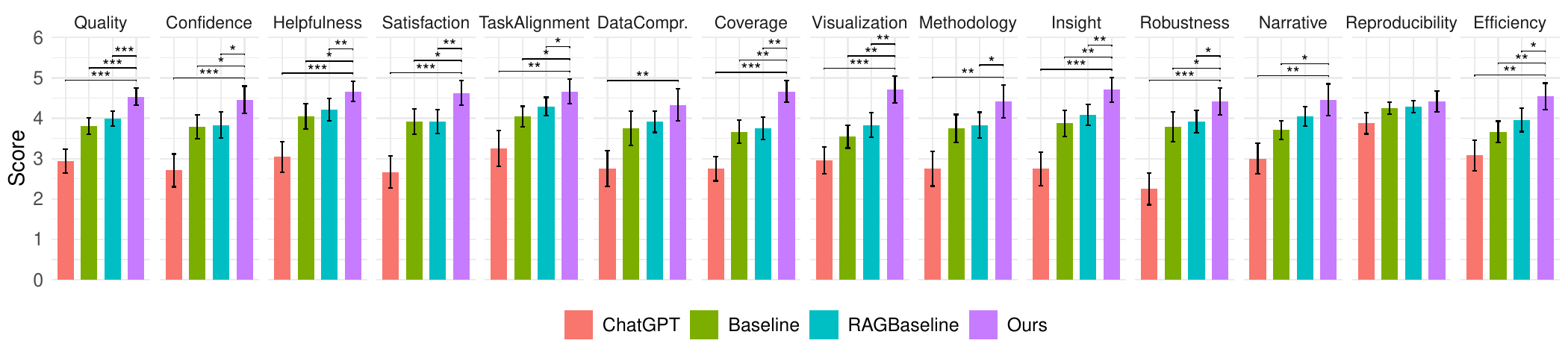}
  \vspace{-22px}
  \caption{Evaluation results of the four notebooks. Error bars show 95\% confidence intervals computed using the Cousineau-Morey method. Asterisks indicate statistical significance ($*$ $p<0.05$, $**$ $p<0.01$, $***$ $p<0.001$) from pairwise Wilcoxon signed-rank tests with Holm-Bonferroni correction. For clarity, only significant comparisons between \textit{Ours} and other notebooks are shown.}
  \label{fig: score_barplot}
  \vspace{-16px}
\end{figure*}

\revised{Since the code output was already run and displayed as in shared notebooks, several participants mentioned that they just read the content. However, most participants still executed the notebook cells themselves. 
Among them, some ran only a few cells to verify the outputs before reading, others executed all cells, and some modified the code for further exploration.
In terms of reading behavior, some participants first navigated the notebook using the ``outline'' function in their environment and jumped to sections of interest, while others read sequentially. Many noted that even though their task was merely to evaluate notebook quality, the way they interacted with the notebooks closely mirrored how they would use them if they were genuinely conducting EDA, such as reusing code directly, checking intermediate outputs, or merely exploring insights. This behavioral alignment supports the representativeness of our participant group.}

The ratings of the four notebooks are shown in \Cref{fig: score_barplot}.
Each bar shows the mean rating across participants.
To handle within-subjects variables, we estimated 95\% confidence intervals using the Cousineau-Morey method \cite{cousineau2005confidence,morey2008confidence}.
Pairwise Wilcoxon signed-rank tests with Holm-Bonferroni correction were applied to assess significance.
We conducted all six pairwise tests, resulting in slightly conservative significance estimates.

\revised{Across the four overall dimensions, \textit{Ours} consistently and significantly outperforms the other three notebooks.  
As overall scores provide a high-level summary, participants often reiterated similar rationales within the corresponding sub-dimensions.
Due to space constraints, we focus directly on the most relevant sub-dimensions, presenting representative quotes and discussing the key mechanisms underlying the observed improvements.
}
To avoid redundancy, we exclude \textit{ChatGPT} due to its uniform inferiority across dimensions.

\begin{itemize}[leftmargin=*,itemsep=2pt,topsep=0pt,parsep=0pt]

    \revised{\item \textbf{\textit{Task Alignment.}} Participants consistently highlighted that \textit{Ours} showed strong task alignment and analytic depth, noting it ``defined objectives early'' and ``adhered closely to prediction goals.'' They felt it ``fully addressed the task of comparing category/subcategory sales and profits, forming a complete chain'', making it the most aligned among the four notebooks (regarding one of the tasks), in contrast to \textit{Baseline} being ``generic'' and \textit{RAGBaseline} sometimes ``lacking connection to decision scenarios.''
    \final{This is further supported by our objective check, which confirmed that \textit{Ours} covered all task-relevant key variables.}
    }

    \revised{\item \textbf{\textit{Visualization.}} \textit{Ours} offered ``diverse and well-matched chart types'' and annotations that made interpretations ``clearer and easier to follow.'' Several noted that it conveyed ``more information without being unreadable.'' They also appreciated that \textit{Ours} avoided readability problems seen elsewhere, such as heatmaps being ``distorted by extreme values.'' (\textit{Baseline}) or messy lineplots under extreme values (\textit{RAGBaseline}). Participants further praised \textit{Ours} for matching visuals to analytic goals, using nested donut charts, faceted scatterplots, and annotated time-series, making it the most effective among the four, even if some charts (e.g., percentage plots or the t-SNE) required more effort to digest.
    The effectiveness stemmed from referencing visualization designs in existing notebooks, which better align with human analytical reasoning.}

    \revised{\item \textbf{\textit{Methodology.}} \textit{Baseline} relied heavily on descriptive statistics and was repeatedly described as ``mentioning ANOVA without proper explanation,'' lacking hypothesis testing or deeper inference. \textit{RAGBaseline} offered a ``coherent workflow'' with some time-series checks and proportioning or bucketing methods, though its statistical validation remained limited. In contrast, \textit{Ours} incorporated richer techniques such as ``bootstrap and regression analysis'' and ``decomposition and autocorrelation diagnostics,'' which participants felt addressed the sub-problems more directly. \textit{Ours} was consistently seen as applying more advanced and appropriate methods than the other notebooks.
    This is because the retrieved content provided relevant analytical examples, enabling the LLM to select and apply more appropriate statistical methods.}

    \revised{\item \textbf{\textit{Insight.}} \textit{Ours} was consistently praised for offering ``deeper and more meaningful insights,'' often supported by ``specific numbers'' such as percentage breakdowns or correlations, and further strengthened by ``actionable recommendations.'' Participants noted that it moved beyond surface observations to examine underlying causes, identify fine-grained scenarios, and provide interpretations ``closely tied to decision-making.'' In comparison, \textit{Baseline} was frequently described as ``shallow,'' offering largely descriptive summaries such as ``organic prices are higher,'' while \textit{RAGBaseline} supplied numerical evidence but remained ``descriptive'' and lacked concrete strategies. \textit{Ours} stood out for producing more useful insights. 
    \final{In addition, our objective correctness check identified a low rate of factual inconsistencies (8/319) across all notebooks generated by \textit{Ours}.}
    This is largely attributed to the agent’s autonomous ability to analyze and adjust its reasoning, leading to more refined and decision-oriented insights.} 

    \item \textbf{\textit{Reproducibility.}} On reproducibility, the study revealed no significant differences for \textit{Ours}. Our intention was to ensure the codes executed normally so participants could assess their quality, making this outcome more a reflection of externally designed scales than of differences. Nevertheless, we retain this sub-dimension to demonstrate that our pipeline produces notebooks that run reliably. 

    \revised{\item \textbf{\textit{Efficiency.}} Participants frequently praised \textit{Ours} for having ``no redundant content,'' offering ``richer insights with less repetition,'' and achieving the ``highest information density.'' \textit{Baseline} was often described as ``too detailed and repetitive.'' While \textit{RAGBaseline} was seen as relatively concise, its insights were sometimes lengthy without focus. \textit{Ours} most effectively condensed analysis while still exploring multiple perspectives and maximizing insight value.
    It results from the agent’s global state management ability, allowing it to maintain context and eliminate redundant analysis.}

\end{itemize}

In summary, the quality of \textit{Ours} has been substantially enhanced by the improved RAG-based pipeline, enabling analyses that are more closely aligned with user intent and more insightful in interpretation. 
\revised{The visualizations are more appropriate to the analysis, both visually and methodologically.}
\final{However, the most frequent negative feedback on \textit{Ours} was that its charts were sometimes unnecessarily complex, which could potentially induce cognitive overload.
We will return to this issue in the discussion section.
We further analyzed the results within each dataset–task subset using participant-level pairwise score differences (see supplementary materials).
Across datasets and tasks, \textit{Ours} consistently achieved higher scores, indicating robust performance.}

%% file: sections/7-discussion.tex
\section{Discussion and Conclusion}
\label{sec: discussion}

\textbf{Constraints on the Methods in Comparative Evaluation.}
\textit{ChatGPT} is designed as an interactive interface, but for a fair comparison with \textit{Ours}, we adopted a fixed prompt template to ensure end-to-end generation, which inevitably constrained its capabilities. Therefore, our experimental conclusions only demonstrate that under this constraint, \textit{Ours} performs better. In addition, since no prior work was directly comparable, the design of \textit{RAGBaseline} combined an existing retrieval scheme originally developed for question answering with our \textit{Baseline}. This may introduce potential unfairness, as the retrieval strategy is not fully tailored to downstream EDA tasks.

\textbf{Infeasibility in Simulating Enterprise Scenarios.}
Enterprises with data analysis needs represent a potential application scenario for our system. 
However, the inaccessibility of internal data and artifacts prevented direct evaluation in this scenario.
In our experimental design, we attempted to approximate enterprise settings through dataset choices, such as using continuously updated datasets and those with multiple versions. However, differences remain compared with real-world enterprise practice, such as the format of notebooks and the nature of specific tasks. 
In future work, we aim to collaborate with enterprises to conduct more realistic evaluations in this scenario, 
\final{where analyses are often deeply coupled with proprietary business logic and contextual knowledge that agentic coding approaches may lack, thereby better highlighting the value of our method.}

\textbf{Possible Influence of Datasets and Tasks.}
While we examined the advantages of \textit{Ours} within the corresponding subsets, we recognize the possibility that interactions may exist between datasets/tasks and our method. 
To account for this, we attempted to fit a mixed-effects linear model that included these interaction terms. 
As noted in \Cref{sec: results}, the estimates were not sufficiently robust to report, given the limited sample size. 
Nonetheless, inspection of the point estimates reveals some deviations from zero, suggesting slight performance variations across datasets or task types, but these coefficients are considerably smaller than the main effect of our method.

\revised{
\textbf{Scalability and Generalizability of the System.}
When the number of notebooks is large and their quality varies, we propose pre-filtering notebooks based on their quality and relevance to the user intent. Previous works have considered multiple criteria, including format \cite{quaranta2022pynblint,pimentel2021understanding}, reproducibility, executability \cite{quaranta2022assessing}, and understandability \cite{ghahfarokhi2024predicting}. Our small-scale tests suggest that SOTA LLMs align more closely with human evaluation, and thus, we consider using LLMs to assess these criteria, as well as the relevance to the user’s intent.
For selected notebooks, static code analysis could be used to identify and exclude those with syntax errors or incomplete code. Since our method does not have strict requirements for markdown content and treats it as plain text, low-quality markdown would not affect the system's ability to operate properly.
We acknowledge that existing notebooks may not always fully address users' concerns or maintain high quality. However, because our retrieval is based on used-column matching rather than semantic similarity, it avoids introducing large amounts of irrelevant content, even when related material is scarce. Small-scale tests on niche tasks further showed that our method performs at least as well as the \textit{Baseline} and \textit{RAGBaseline}. \final{When no relevant notebooks are available or the dataset is non-tabular format, the baseline generator can still produce reasonable results, though this is not the primary focus of our work.}
}

\textbf{Possible Over-reliance on Existing Corpora.}
Feedback from the user study indicated that notebooks produced by \textit{Ours} sometimes included overly complex charts.
This was likely because highly upvoted Kaggle notebooks may use elaborate designs to attract attention, although such cases were infrequent.
\final{Since some participants appreciated high-information-density visualizations, we plan to adapt our generation strategy in future work based on assessments of visualization complexity and users’ visualization preferences}.

\textbf{Execution Efficiency.}
Compared with \textit{Baseline}, \textit{Ours} requires additional time in the retrieval stage. Since components are independent, we employ a parallel strategy that keeps the processing time for about twenty notebooks (roughly 300 components) within three minutes. The main bottlenecks lie in leveraging the VLM for insight extraction and in code debugging. 
Given that generating a complete EDA notebook (with twelve sub-tasks) typically takes around ten minutes, the retrieval cost remains within an acceptable range.
\final{Given that generating a complete EDA notebook (with twelve sub-tasks) typically takes around ten minutes, and the system is designed as an offline assistant, the retrieval overhead remains acceptable.}

\textbf{Insight Generation.}
In the Component Enhancement stage (\Cref{sec: enhancement}), we adopted a strategy of first extracting insights from visualizations and then refining them with statistical code, which helps mitigate hallucinations from VLMs. 
\revised{Theoretically, this approach better leverages the strengths of visualizations in revealing patterns; however, our current evaluation is limited to a small-scale manual review and is not yet comprehensive. We acknowledge that this pipeline may introduce a potential confirmation bias, as the statistical verification is performed based on the insights initially proposed by the VLM. In future work, we plan to address this limitation by exploring alternative designs, such as generating and comparing multiple competing hypotheses from the same visualization or explicitly incorporating falsification-oriented statistical tests. We also aim to conduct a more systematic evaluation of how this two-stage reasoning process influences both reliability and interpretability.}


